\documentclass[pre, twocolumn]{revtex4}
\usepackage{amsmath}
\usepackage{braket}
\usepackage[dvipdfmx]{color}
\usepackage[dvipdfmx]{graphicx}

\begin{document}
\title{Condition for emergence of the Floquet-Gibbs state in periodically driven open systems}
\author{Tatsuhiko Shirai}
\email{shirai@spin.phys.s.u-tokyo.ac.jp}
\author{Takashi Mori}
\email{mori@spin.phys.s.u-tokyo.ac.jp}
\author{Seiji Miyashita}
\email{miya@spin.phys.s.u-tokyo.ac.jp}
\affiliation{%
Department of Physics, Graduate School of Science,
The University of Tokyo, 7-3-1 Hongo, Bunkyo-Ku, Tokyo 113-8656, Japan
}

\begin{abstract}
We study probability distribution of a steady state of a periodically driven system coupled to a thermal bath by using a quantum master equation in the weak coupling limit.
It is proved that, even when the external field is strong,
the probability distribution is independent of the detailed nature of the thermal bath under the following conditions:
(i) the Hamiltonian of the relevant system is bounded and the period of the driving field is short, (ii) the Hamiltonians for the driving field at different times commute, and (iii) the Hamiltonians of the driving field and of the interaction between the relevant system and the thermal bath commute.
It is shown that the steady state is described by the Gibbs distribution of the Floquet states of the relevant system at the temperature of the thermal bath.
\end{abstract}

\maketitle

Quantum systems under a periodic driving field exhibit many interesting phenomena.
For instance, when an atom resonates with an electromagnetic field, Rabi oscillation appears.
For specific values of the field amplitude and frequency, quantum tunneling is totally suppressed,
which is known as the dynamical localization~\cite{dunlap1986dynamic} and the coherent destruction of tunneling (CDT)~\cite{grossmann1991coherent,grifoni1998driven}.
The time evolution of a periodically driven quantum system is described by the Floquet theorem.
These various characteristic phenomena of the driven systems are understood from the viewpoint of eigenvalues of the Floquet operator~\cite{sambe1973steady, grossmann1991coherent, llorente1992tunneling}.

It is an unsolved issue how the steady state of the system subject to the periodic driving and the dissipation is characterized.
In the weak-coupling limit between the system of interest and the thermal bath, it is shown that the dynamics of the relevant system can be described as a process of transitions among the Floquet states~\cite{blumel1991dynamical,kohler1997floquet,breuer2000quasistationary,langemeyer2014energy,Dehghani2014dissipative}.
Therefore, the driven steady state is characterized by the probability distribution among the Floquet states.
In the case without driving field, the system relaxes to the Gibbs distribution, which is independent of the details of the thermal bath except for its temperature.
The detailed balance condition for the transition probabilities between the energy eigenstates ensures this insensitivity to the structures of the thermal bath.
On the other hand, under the driving field, the transition probabilities between the Floquet states do not satisfy the detailed balance condition in general and the probability distribution among the Floquet states in a driven steady state may depend on the thermal bath~\cite{kohn2001periodic}.
 
In this Rapid Communication it is pointed out that under some conditions the driven steady state of the relevant system is described by the Gibbs distribution of the Floquet states,
which is called the Floquet-Gibbs state here, and hence the driven steady state is independent of the details of the thermal bath.
It was noted that the Floquet-Gibbs state is realized when the time dependence of the total system is completely removed by using a unitary transformation on the relevant system~\cite{iadecola2013generalized}.
In the case of a linearly driven harmonic oscillator, the total Hamiltonian after a unitary transformation~\cite{husimi1953miscellanea} is approximately time independent~\cite{breuer2000quasistationary}.
The remaining time dependence does not affect the dynamics of the relevant system within the lowest order of system-bath coupling.
However, the scenario to realize the Floquet-Gibbs states strongly depends on the forms of the Hamiltonians ruling their dynamics.
Here, we consider two situations which are not restricted to specific models.
One is the driven system in the linear response regime.
It was pointed out that the driven steady state is almost independent of the details of the thermal bath in earlier works, see for example~\cite{hone2009statistical}.
The other situation is that the driving field has a very high frequency and very large amplitude beyond the linear-response regime.
We show that the condition of the high-frequency driving field is not sufficient for the Floquet-Gibbs state to realize,
and thus the Floquet-Gibbs state is a rather restrictive feature of a driven open system.
Later, we give two more conditions (ii) and (iii) for the realization.

We study the dynamics by using the quantum master equation for the periodically driven system.
Here, we overview it to explain the notation for the following discussion.
The Hamiltonian of the total system is given by $H(t)=H_{\rm S}(t)+H_{\rm B}+\lambda H_{\rm I}$,
where $H_{\rm S}(t)$ and $H_{\rm B}$ are the Hamiltonians of the relevant system and the thermal bath, respectively,
and $H_{\rm I}$ represents the interaction between them.
The coupling strength of the interaction Hamiltonian is denoted by $\lambda$.
The Hamiltonian of the relevant system is given by 
\begin{equation}
H_{\rm S}(t)=H_0+ A H_{\rm ex} (t),
\end{equation}
where $H_{\rm ex} (t)$ stands for the driving field with period $T$ and $A$ denotes the amplitude of the driving field.
Without loss of generality, we assume that $\int_0^T H_{\rm ex} (t) dt=0$.
The interaction Hamiltonian $H_{\rm I}$ is in the form $\sum_{\mu}X_{\mu} Y_{\mu}$ where $X_{\mu}$ and $Y_{\mu}$ are the operators on the relevant system and the thermal bath, respectively.

In the weak coupling limit ($\lambda \rightarrow 0$), the master equation of the Born-Markov type for the reduced density operator $\rho_{\rm S}(t)$ is given by~\cite{nakajima1958quantum, zwanzig1960ensemble, kubo1991statistical, petruccione2002theory}
\footnote{The validity of this master equation is discussed in the appendix A in~\cite{hone2009statistical}.}
\begin{align}
&\frac{d}{dt}\rho_{\rm S}(t)=-\frac{i}{\hbar} [H_{\rm S}(t), \rho_{\rm S} (t)] \nonumber\\
&-\frac{\lambda^2}{\hbar^2} \sum_{\mu, \nu} \int_0^{\infty} d\tau 
 \Big\{ \langle Y_{\mu}(\tau ) Y_{\nu} \rangle_{\beta} [ X_\mu, X_{\nu}(t, t-\tau) \rho_{\rm S}(t) ]\nonumber\\
& \qquad\qquad\quad -\langle Y_{\nu} Y_{\mu} (\tau ) \rangle_{\beta} [ X_{\mu}, \rho_{\rm S}(t) X_{\nu}(t, t-\tau ) ] \Big\}. \label{master_eq}
\end{align}
Here, 
$X_{\mu}(t', t)=U(t', t) X_{\mu} U^{\dagger} (t', t)$ where $U(t', t)=\exp_{\leftarrow} \left( - \frac{i}{\hbar} \int_{t}^{t'} H_{\rm S}(\tau ) d\tau \right) $  and $Y_{\mu}(t)=e^{i H_{\rm B} t /\hbar } Y_{\mu} e^{ - i H_{\rm B} t/\hbar }$.
The subscript arrow stands for the usual time-ordering operator.
The angular bracket $\langle \cdots \rangle_{\beta} $ is the average over the canonical distribution of $H_{\rm B}$ at the inverse temperature $\beta$.
The dependence on the thermal bath fully comes from the time correlation functions $\langle Y_{\mu}(\tau ) Y_{\nu} \rangle_{\beta}$.
We introduce its Fourier transform, 
\begin{equation}
S_{\mu, \nu} (\epsilon )=\int_{-\infty}^{\infty} \langle Y_{\mu}(t) Y_{\nu} \rangle_{\beta} e^{\frac{i}{\hbar} \epsilon t} \frac{dt}{2\pi}.
\end{equation}
The form of the function $S_{\mu, \nu} (\epsilon )$ depends on the bath operators $H_{\rm B}$ and $\{ Y_{\mu} \}$.
It satisfies the Kubo-Martin-Schwinger (KMS) relation~\cite{kubo1957statistical},
$S_{\mu, \nu}(\epsilon )=S_{\nu, \mu} (-\epsilon ) e^{\beta \epsilon}$,
which holds independently of the details of the thermal bath.
This relation plays a role to thermalize the relevant system in the case without the driving external field, $H_{\rm ex} (t)=0$.
In other words, the KMS relation leads to the detailed balance condition at least in O$(\lambda^2)$,
\begin{equation}
P_{p \rightarrow q} e^{-\beta E_p} =P_{q \rightarrow p} e^{-\beta E_q}, \label{detailed_balance}
\end{equation}
where $E_p$ is the eigenenergy of $H_0$.

Because Eq.~(\ref{master_eq}) is described by the Hamiltonian with a periodic field,
it is convenient to write the equation in the basis of the Floquet states of the relevant system,
$\ket{u_p (t)}=U(t, 0) e^{i \epsilon_p t/\hbar}\ket{u_p (0)} $,
where $\ket{u_p (0)}$ is the eigenstate of the time evolution operator over one period,
$U(T, 0) \ket{u_p (0)}=e^{-i \epsilon_p T/\hbar } \ket{u_p (0)}$.
Here, the quesienergy $\epsilon_p$ is chosen to lie in the region $-\pi \hbar /T \leq  \epsilon_p < \pi \hbar/T$.
The Floquet state thus defined is periodic, $\ket{u_p (t)}=\ket{u_p (t+T)}$.
Matrix elements of the density operator in the basis of the Floquet states are denoted by $\rho_{{\rm S},pq}(t)=\bra{u_{p}(t)} \rho_{\rm S}(t) \ket{u_q (t)}$.
We assume the non-resonance condition; if $(\epsilon_p -\epsilon_q) -2\pi m \hbar/T =(\epsilon_{p'} -\epsilon_{q'}) -2\pi {m'}\hbar/T $ with some integers $m$ and $m'$ for $p \neq q$, then
$p=p', q=q',$ and $m=m'$.
Then the time evolution of the diagonal elements and the off-diagonal elements are decoupled if we neglect the oscillating terms (the rotating wave approximation).
The diagonal elements obey the equation,
\begin{equation}
\frac{d}{dt}\rho_{{\rm S},pp}(t)=\sum_{q (\neq p)} ( P_{q \rightarrow p} \rho_{{\rm S},qq} (t)- P_{p \rightarrow q} \rho_{{\rm S},pp} (t) ),\label{diagonal_eq}
\end{equation}
where $P_{p \rightarrow q}$ is the transition probability from $\ket{u_p (t)}$ to $\ket{u_q (t)}$.
Off-diagonal elements obey $d \rho_{{\rm S},pq}(t) /dt=-\Gamma_{pq} \rho_{{\rm S},pq}(t)$,
where $\Gamma_{pq}$ is a complex quantity and Re$\Gamma_{pq} >0$.
Hence, the off-diagonal elements go to zero in the long-time limit and the driven steady state is described by a diagonal matrix,
\begin{equation}
\rho_{\rm s.s.}(t)= \sum_{p} P_{p} \ket{u_{p} (t)} \bra{u_{p}(t)},
\end{equation}
where $P_{p}=\lim_{t \rightarrow \infty} \rho_{{\rm S},pp}(t)$.
Since the Floquet state is periodic, the steady state also has the same periodicity, $\rho_{\rm s.s.}(t)=\rho_{\rm s.s.}(t+T)$~\cite{blumel1991dynamical,breuer2000quasistationary}.

Now, we study distribution of the steady state $\{ P_{p} \}$
\footnote{Here, we focus on a time-periodic asymptotic state up to the leading order of $\lambda$,
which is not affected by the rotating wave approximation
when the system Hamiltonian does not possess any degeneracies in its quasienergy eigenspectrum~\cite{mori2008dynamics,thingna2013reduced,yuge2015perturbative}.}.
For this, we need concrete forms of the transition probabilities which are given by
\begin{equation}
P_{p \rightarrow q}=\frac{2\pi \lambda^2}{\hbar} \sum_{m=-\infty}^{\infty} P_{p \rightarrow q, m}, \label{diagonal_master}
\end{equation}
where 
\begin{equation}
P_{p \rightarrow q, n}=\sum_{\mu, \nu} X_{\mu, pqn}X_{\nu, qp-n} S_{\mu, \nu}\left( \epsilon_p-\epsilon_q+\frac{2\pi n \hbar}{T} \right)
\end{equation}
and
\begin{equation}
X_{\mu, pqn}=\int_0^T \bra{u_p (t)} X_\mu \ket{u_q (t)} e^{ -2\pi i n \frac{t}{T}} \frac{dt}{T}. \label{X}
\end{equation}

Under the periodic driving,
the relation similar to the detailed balance condition does not hold in general for the transition probabilities among the Floquet states due to the sum of $m$ in Eq.~(\ref{diagonal_master})~\cite{kohn2001periodic}.
In general, the driven steady state depends on the details of the thermal bath $S_{\mu, \nu}(\epsilon)$.

However if only one term of $m$ in Eq.~(\ref{diagonal_master}) denoted by $m_{pq}$ gives a dominant contribution, and furthermore $m_{pq}$ is given by $m_p-m_q$,
the detailed balance condition is satisfied,  
\begin{equation}
P_{p \rightarrow q} e^{-\beta (\epsilon_p +2 \pi m_p \hbar/T )}=P_{q \rightarrow p} e^{-\beta (\epsilon_q +2 \pi m_q \hbar/T )}.\label{transition}
\end{equation}
The introduction of the integer $m_p$ for each Floquet state allows us to assign ``energy'' $E_p=\epsilon_p+2\pi m_p \hbar /T$,
and construct the ``Floquet Hamiltonian'' $H_{\rm F}(t)=\sum_p E_p \ket{u_p (t)} \bra{u_p (t)}$.
The driven steady state is then expressed by the Floquet-Gibbs state,
$\rho_{\rm s.s.}(t) \propto e^{-\beta H_{\rm F}(t)}$,
which does not depend on the structures of the thermal bath.
We shall show that the above situation occurs in some physically relevant cases.

First, we study the linear response regime where the strength of the driving field is weak $A \ll 1$,
and confirm that, up to the first order of $A$, the probability distribution of the steady state is independent of the details of the thermal bath.
Since the system is weakly driven, the Floquet state is expanded by the eigenstate of $H_0$, $H_0 \ket{\phi_p}= E_p \ket{\phi_p}$.
By using a perturbation method~\cite{sambe1973steady}, the Floquet state is given by
\begin{align}
&\ket{u_{p}(t)}e^{2\pi i m_p^0 \frac{t}{ T}} = ( e^{-\frac{i}{\hbar} A \int_0^t  \bra{\phi_p} H_{\rm ex}(\tau ) \ket{\phi_p} d\tau } +{\rm O}(A^2)) \ket{\phi_p} \nonumber\\
& - \sum_{q (\neq p)} \left( A \sum_{m (\neq 0)}^{\infty} \frac{ \bra{\phi_q} \tilde{H}_{m} \ket{\phi_p} e^{-2\pi i m \frac{t}{T}} }{ E_q-E_p-2\pi m \hbar/T} +{\rm O}(A^2) \right) \ket{\phi_q},
\end{align}
where $\tilde{H}_m=\int_0^T H_{\rm ex}(t) e^{2\pi i m t/T} dt /T$,
and $m_p^0$ is defined by $\epsilon_p^0=E_p -2\pi m^0_p \hbar/T$ to satisfy $-\pi \hbar/T \leq \epsilon_p^0 < \pi \hbar /T$.
We have assumed the condition for the perturbation method, $|E_p -E_q+2\pi n\hbar/T | \gg A \max_t |\bra{\phi_p} H_{\rm ex} (t) \ket{\phi_q}|$ for any set of $(p, q, n)$ except $p=q$ and $n=0$.
We then obtain $X_{\mu, pqn}$ in Eq.~(\ref{X}) for $n=m_p^0-m_q^0$,
\begin{align}
X_{\mu, pqn}=& \Big[ 1+A \sum_{m(\neq 0)}^{\infty} \frac{\bra{\phi_q} \tilde{H}_{m} \ket{\phi_q}-\bra{\phi_p} \tilde{H}_{m} \ket{\phi_p} }{2\pi m \hbar/T} \Big] \nonumber\\
&\times \bra{\phi_p} X_{\mu} \ket{\phi_q} +{\rm O} (A^2 ),
\end{align}
and for $n \neq m_p^0-m_q^0$, $X_{\mu, pqn}={\rm O}(A)$.
The transition probability $P_{p \rightarrow q}$ is given by
\begin{align}
&P_{p \rightarrow q}=\frac{2\pi \lambda^2}{\hbar} \sum_{\mu, \nu} \bra{\phi_p}X_{\mu} \ket{\phi_q} \bra{\phi_q} X_{\nu} \ket{\phi_p} \nonumber\\
&\times S_{\mu, \nu}( (\epsilon_p +2\pi m_p^0 \hbar /T)- (\epsilon_q +2\pi m_q^0 \hbar /T))+{\rm O}(A^2).
\end{align}
An integer $m_p^0$ is assigned for each Floquet state which indicates that the steady state is independent of the details of the thermal bath 
\footnote{There is some freedom to add an integer to each $m_p$ because the origin of $m_p$ is not important.}.
The KMS relation leads to the detailed balance condition,
\begin{equation}
P_{p \rightarrow q} e^{-\beta \left( \epsilon_p +\frac{2\pi m_p^0 \hbar}{T} \right)}= P_{q \rightarrow p} e^{-\beta \left( \epsilon_q + \frac{2\pi m_q^0 \hbar}{T} \right) }+{\rm O}(A^2).
\end{equation}
If the ergodicity is satisfied in the limit $A \rightarrow 0$,
the driven steady state is given by $\rho_{\rm s.s}(t) \propto e^{-\beta H_{\rm F}(t)}+{\rm O}(A^2 )$, where $H_{\rm F}(t)=\sum_{p} ( \epsilon_p+2\pi m_p^0 \hbar /T ) \ket{u_{p} (t)}\bra{u_{p} (t)}$.
From the fact that $\epsilon_p +2\pi m_p^0 \hbar /T=E_p +{\rm O} ( A^2 )$,
the driven steady state can be described by $\rho_{\rm s.s}(t) \propto e^{-\beta E_p} \ket{u_p (t)} \bra{u_p (t)}+{\rm O}(A^2 )$.
Although the distribution of the driven steady state $\{ P_p \}$ is the same as the equilibrium state in the order of $A$,
the steady state is different from the equilibrium state 
because the Floquet state changes in this order.

Next, we consider the case of large $A$ beyond the linear response regime.
For this purpose, we impose the following conditions:
(i) the period of the driving field is short, $\delta \equiv \| H_0 \| T/\hbar \ll 1$,
and (ii) for any pair of $t_1$ and $t_2$, $[H_{\rm ex}(t_1), H_{\rm ex}(t_2)]=0$.
It is noted that, because of the condition (i), the following argument is relevant for a relatively small system, in which $\| H_0 \|$ is not so large.

These conditions lead us to introduce a time-averaged Hamiltonian in a rotating frame.
The state in the rotating frame $\ket{\Psi^R (t)} $ is related to a state in the static frame $\ket{\Psi (t)}$ by a unitary transformation,
$\ket{\Psi^R (t)}=e^{i \int_0^t A H_{\rm ex} (\tau) d\tau /\hbar} \ket{\Psi (t)}$.
In this frame, the time evolution is governed by the Hamiltonian $H_{\rm T}^R(t)=H_{\rm S}^R(t)+\lambda H_{\rm I}^R(t)+H_{\rm B}$
where $H_{\rm S}^R (t)$ and $H_{\rm I}^R (t)$ are given by
\begin{align}
H_{\rm S}^R (t)=&e^{\frac{i}{\hbar} \int_0^t A H_{\rm ex} (\tau) d\tau} H_0 e^{-\frac{i}{\hbar} \int_0^t A H_{\rm ex} (\tau) d\tau},\nonumber\\
H_{\rm I}^R (t)=&e^{\frac{i}{\hbar} \int_0^t A H_{\rm ex} (\tau) d\tau} H_{\rm I} e^{-\frac{i}{\hbar} \int_0^t A H_{\rm ex} (\tau) d\tau}.\label{interaction}
\end{align}
The condition (ii) ensures the periodicity of these Hamiltonians,
and we expand them into the Fourier series, $H_{\alpha=\{{\rm S, I}\}}^R (t)=\sum_{m} \tilde{H}_{\alpha=\{{\rm S, I}\}, m}^R e^{-2\pi i m t/T}$.
Because the norm of $H_{\alpha=\{{\rm S, I}\}}^R (t)$ does not depend on $A$, $H_{\alpha=\{{\rm S, I}\}}^R (t)$ consists of the rapidly oscillating terms with finite amplitude.
We may have an intuitive picture that $H_{\alpha=\{{\rm S, I}\}}^R (t)$ is replaced by the time-averaged Hamiltonian over a period, $\tilde{H}_{\alpha=\{{\rm S, I}\}, 0}^R=\int_0^T H_{\alpha=\{{\rm S, I}\}}^R (t) dt/T$
\footnote{
This procedure for moving to a rotating frame and then neglecting rapidly oscillating terms can be regarded as a rotating wave approximation.
See the applications to systems under the driving field with a high frequency, e.g., CDT~\cite{llorente1992tunneling,kayanuma1994role,ashhab2007two,creffield2010coherent,gomez2011charge}, many-body systems~\cite{eckardt2005superfluid,bastidas2012nonequilibrium}, and so on~\cite{bukov2014universal}.}.
Since the total Hamiltonian $H_{\rm T}^R (t)$ is then replaced by the averaged one, $\tilde{H}_{{\rm S}, 0}^R+\lambda \tilde{H}_{{\rm I}, 0}^R+H_{\rm B}$, the steady state of the relevant system seems to be described by the Gibbs state of $\tilde{H}_{{\rm S}, 0}^R$.

However, the detailed balance condition does not generally hold even when these two conditions (i) and (ii) are satisfied.
That is, the time dependence in Eq.~(\ref{interaction}) nonperturbatively changes the steady state in general.
Nonetheless, we find that if we assume the additional condition (iii), $[H_{\rm ex} (t), H_{\rm I}]=0$,
only the $m=0$ term in Eq.~(\ref{diagonal_master}) gives a dominant contribution and hence the detailed balance condition is satisfied.

Now, we show that the relevant system reaches the Floquet-Gibbs state under the three conditions (i), (ii), and (iii).
For this purpose, we introduce the Floquet state in the rotating frame $\ket{u^R_{p} (t)}$ and its quasienergy $\epsilon^R_{p}$ associated with $H_{\rm S}^R(t)$ which is related to those in the static frame by
\begin{equation}
\ket{u^R_{p} (t)}=e^{\frac{i}{\hbar} \int_0^t A H_{\rm ex} (\tau) d\tau} \ket{u_{p} (t)}, \quad \epsilon^R_{p}=\epsilon_{p}.\label{rotating}
\end{equation}
Due to the condition (iii), $ \ket{u_{p} (t)}$ is replaced by $\ket{u^R_{p} (t)}$ in $X_{\mu, lmn}$(Eq.~(\ref{X})).
By definition,
\begin{equation}
\ket{u_p^R (t)}=\exp_{\leftarrow} \left( -\frac{i}{\hbar} \int_0^t (H_{\rm S}^R(\tau ) -\epsilon_p^R ) d\tau \right) \ket{u_p^R (0)},
\end{equation}
which is expanded into the power series of $\delta$ (Magnus expansion~\cite{blanes2009magnus}),
\begin{align}
\ket{u_p^R (t)}=&\Big[ 1-\sum_{m(\neq 0)}  \frac{\tilde{H}_{{\rm S},m}^R(1-e^{-2\pi i m \frac{t}{T}})}{2\pi m\hbar} T \nonumber\\
&+{\rm O}(\delta^2 ) \Big] \ket{u_p (0)}.
\end{align}
Here, $\delta$ is independent of $A$, which reflects the fact that the strength of the oscillating terms in the rotating frame is order of $\| H_0 \|$ independent of $A$.
We then obtain $X_{\mu, pqn}$, for $n=0$,
\begin{equation}
X_{\mu, pq0}= \bra{u_p (0)} X_{\mu} \ket{u_q(0)}+{\rm O} ( \delta ),
\end{equation}
and for $n \neq 0$,
\begin{equation}
X_{\mu, pqn}= \frac{T}{2\pi n \hbar} \bra{u_p (0)} [\tilde{H}_{{\rm S},-n}^R, X_{\mu} ] \ket{u_q (0)} +{\rm O}( \delta^2 ).
\end{equation}
The transition probability is then given by
\begin{align}
P_{p \rightarrow q}=& \frac{2 \pi \lambda^2}{\hbar}\sum_{\mu, \nu} \Big[ \bra{u_p (0)} X_{\mu} \ket{u_q (0)} \bra{u_q (0)} X_{\nu} \ket{u_p (0)}\nonumber\\
&+{\rm O}(\delta ) \Big] S_{\mu, \nu}( \epsilon_p  - \epsilon_q )+{\rm O}(\delta^2 ).
\end{align}
For each Floquet state $\ket{u_p (t)}$, an integer $m_p=0$ is assigned.
The KMS relation leads to the detailed balance condition, $P_{p \rightarrow q} e^{-\beta \epsilon_p }= P_{q \rightarrow p} e^{-\beta \epsilon_q }+{\rm O}(\delta^2)$.
If the ergodicity is satisfied in the limit $T \rightarrow 0$ with $AT$ held fixed,
the driven steady state is given by
\begin{equation}
\rho_{\rm s.s.}(t) = e^{-\beta H_{\rm F} (t)}/Z +{\rm O}(\delta^2 ),\label{Gibbs}
\end{equation}
where $H_{\rm F} (t)=\sum_{p} \epsilon_p \ket{u_{p} (t)}\bra{u_{p} (t)}$ and $Z$ is a normalization constant.
In this way, under the conditions (i), (ii), and (iii), the driven steady state is described by the Floquet-Gibbs state.

This result is consistent with the idea of the time-averaged Hamiltonian.
Since $H_{\rm S}^R (t)=\tilde{H}_{{\rm S}, 0}^R +{\rm O}(\delta)$ due to the Magnus expansion,
the driven steady state in the rotating frame is given by the equilibrium state of the averaged Hamiltonian, $e^{-\beta \tilde{H}_{{\rm S}, 0}^R} / {\rm Tr}e^{-\beta \tilde{H}_{{\rm S}, 0}^R}$.

\begin{figure}[t]
\begin{tabular}{cc}
&\hspace{6mm}\includegraphics[width=40mm]{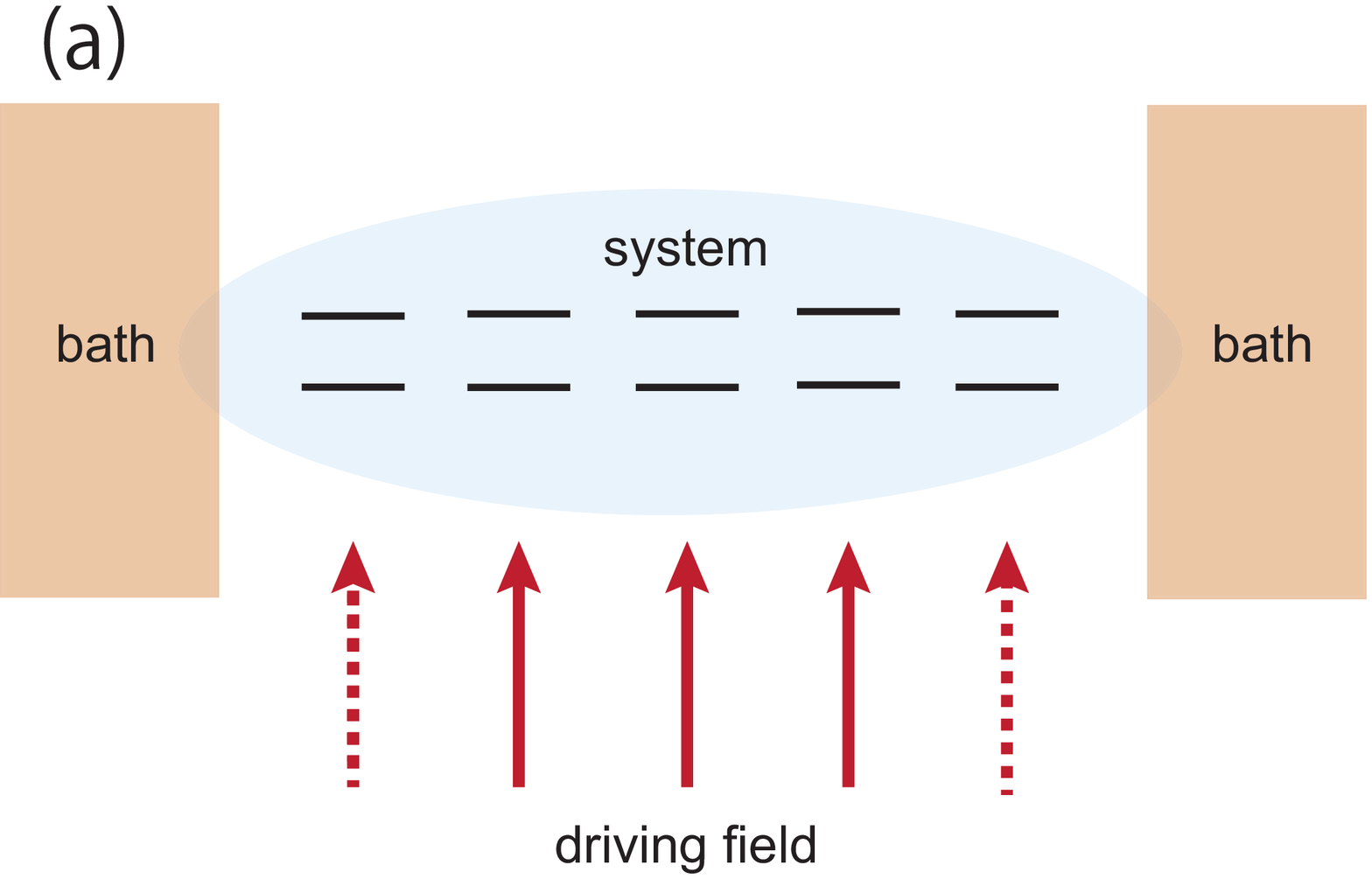}\\
&\includegraphics[width=40mm, bb=0 0 500 616]{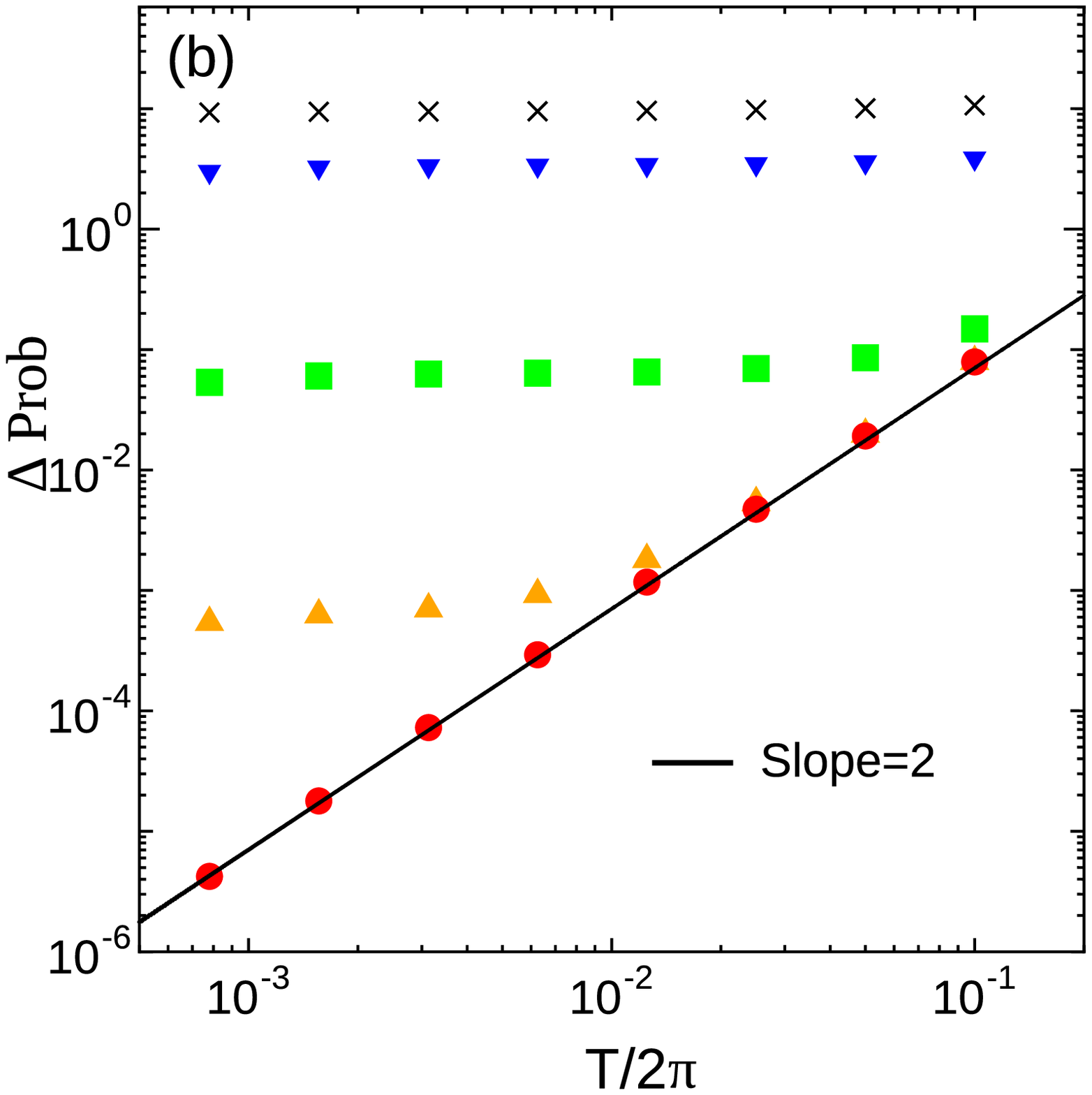}
\includegraphics[width=40mm, bb=0 0 500 616]{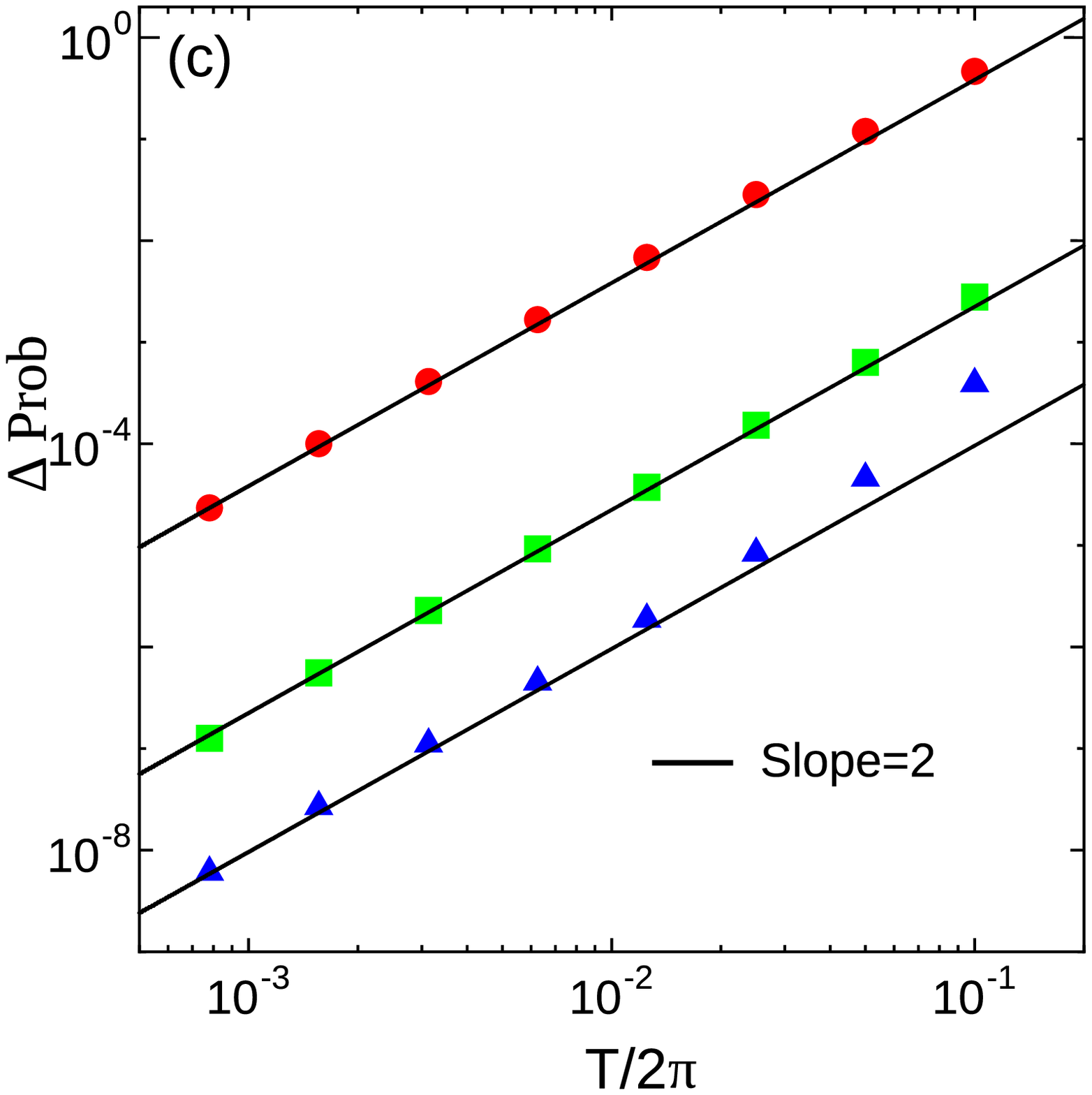}
\end{tabular}
\caption{
(color online). (a) Schematic illustration of our model. (b) $\Delta$Prob versus the period of the driving field $T$ where the relevant system is coupled to the thermal baths via $X_{1}=\sum_{k=\{ x, y, z \}} \alpha^{k} S_1^{k}$ and $X_{5}=\sum_{k=\{ x, y, z \}} \alpha^{k} S_5^{k}$ for various realizations of $\alpha^y / \alpha^x=\{ 0$ (filled circle), $0.01$ (up-triangle), $0.1$ (square), $1$ (down-triangle), $\infty$ (cross) $\}$, and $\alpha^z=0$.
When $[H^{\rm ex}(t), H_{\rm I}]=0$ (filled circle), the steady state approaches the Floquet-Gibbs state.
The larger $\alpha^y / \alpha^x$ is, the larger $\Delta$Prob is.
(c) Like (b), but the driven spins and the spins coupling to the thermal bath are separated.
We study the set $(\alpha^x, \alpha^y, \alpha^z)=\{ (0.1, 0.1, 1)$ (filled circle), $(0.1, 1, 0.1)$ (triangle), and $(1, 0.1, 0.1)$ (square) $\}$.
The steady state approaches the Floquet-Gibbs state independently of the set.
}
\label{sx}
\end{figure}

We demonstrate the above property in a spin system (see in Fig.~\ref{sx} (a)) whose Hamiltonian is given by
\begin{equation}
H_{0}=\sum_{i=1}^5 \left( \omega S_i^z + h  S_i^x \right) -g \sum_{i=1}^4 (S_i^x S_{i+1}^x +S_i^y S_{i+1}^y),\label{model}
\end{equation}
where $\omega=1$, $g=0.2$, and $h=0.01$.
Spins at the edge $i=1$ and $i=5$ are coupled to the independent thermal baths, respectively, 
which are at the same temperature, $\beta=10$, and have the same form of $S_{\mu, \mu}(\epsilon)$, $S_{1,1}(\epsilon)=S_{5, 5}(\epsilon) =\omega_{\rm C}^2/((\epsilon/\hbar)^2+\omega_{\rm C}^2)$.
Here, $\omega_{\rm C}$ is a cutoff frequency, $\omega_{\rm C}=5000$, which is larger than the frequency of the driving field.
The relevant system couples to the thermal bath via $X_{1}=\sum_{k=\{ x, y, z \} } \alpha^k S_1^k$ and $X_{5}=\sum_{k=\{ x, y, z \} } \alpha^k S_5^k$, respectively.
We study following two cases of driving field:
\begin{equation}
\left\{
\begin{aligned}
&A H_{\rm ex}^{(1)}(t)= A \sum_{i=1}^5 S_i^x \cos (2\pi t/T ), A=4\pi/ T\\
&A H_{\rm ex}^{(2)}(t)= A \sum_{i=2}^4 S_i^x \cos (2\pi t/T ), A=4\pi/ T.
\end{aligned} 
\right.
\end{equation}
In both cases, condition (ii) is satisfied.
In the first case, condition (iii) is satisfied only when $\alpha^x \neq 0, \alpha^y=\alpha^z=0$, while in the second case, condition (iii) is always satisfied because the spins under driving and the spins in contact with the thermal baths are separated.

As a measure of the difference between the driven steady state and the Floquet-Gibbs state,
we calculate the deviation,
\begin{equation}
\Delta {\rm Prob}=\sum_p \left| {P}_p -e^{-\beta \epsilon_p}/Z \right|.
\end{equation}

In Fig.~\ref{sx}(b), we study the first case for various values of the ratio $\alpha^y /\alpha^x$.
Here, we set $\alpha^z=0$.
For $\alpha^y=0$, then $[H_{\rm ex} (t), H_{\rm I}]=0$,
the driven steady state approaches the Floquet-Gibbs state as the period of the driving field is short
and $\Delta{\rm Prob}$ is proportional to $T^2$ (filled circle in figure), which agrees with the result, Eq.~(\ref{Gibbs}).
When the ratio $\alpha^y /\alpha^x$ increases, $\Delta$Prob deviates from the $T^2$ dependence due to the violation of condition (iii).

In contrast, in the second case depicted
in Fig.~\ref{sx}(c), the driven steady state approaches the Floquet-Gibbs state with the $T^2$ dependencies independently of the couplings.

In summary, we gave the sufficient conditions under which a driven steady state can be described as the Floquet-Gibbs state
even when the system goes beyond the linear response regime.
This work reveals the applicability of the Floquet-Gibbs state to a steady state of a driven system surrounded by a thermal bath.
It has been argued that the Floquet Hamiltonian $H_{\rm F}(t)$ realizes new phases not accessible without driving field in a cold atomic gas well isolated from the environment~\cite{eckardt2005superfluid,creffield2006tuning}, a semiconductor quantum well~\cite{lindner2011floquet}, and a quantum optical system~\cite{bastidas2012nonequilibrium} ( for a review, see~\cite{bukov2014universal}).
It is naively expected that the averaged Hamiltonian $\tilde{H}_{{\rm S}, 0}^R$ can be used for description of a system with a fast oscillating driving field regardless of whether the system is coupled to a thermal bath or not, but this argument is incorrect.
When the system under a high-frequency driving field is coupled to a thermal bath,
the steady state is not necessarily described by the Floquet-Gibbs state.
Conditions (ii) and (iii) put a constraint on the applicability of the Floquet-Gibbs state.
In particular, condition (iii) is not expected to be satisfied in general.
In that case, the Floquet-Gibbs state is useless as demonstrated in Fig.~\ref{sx} (b).
The Floquet-Gibbs state is relevant for the description of the driven steady state
when the driven spins are different from the spins directly coupling to thermal baths, as in Fig.~\ref{sx} (c).
These results can be implemented in a small system under a periodical driving field coupled to a dissipative environment.
For instance, the theory is applicable to a system consisting of quantum dots under microwaves which are surrounded by two-dimensional electron gas~\cite{van2002electron}.

It is interesting to extend this study to less restrictive situations.
It has been reported that even if the above conditions are not satisfied, there is a situation in which the concept of an effective temperature is useful for the description of the steady state~\cite{verso2010dissipation}.
Besides, our theory is applicable only to a relatively small system due to condition (i).
In recent studies, it is pointed out that a periodic driving field plays a role to change the macroscopic properties of materials:
nonequilibrium phase transitions associated with CDT under the periodic driving~\cite{eckardt2005superfluid,creffield2006tuning,shirai2014novel},
a topological state induced by applying an ac electric field or an ac magnetic field~\cite{lindner2011floquet,delplace2013merging} and by exciting phonon modes~\cite{iadecola2013materials,iadecola2014topological},
and so on~\cite{tsuji2008correlated,bastidas2012nonequilibrium,vorberg2013generalized}.
In macroscopic systems, there appears an energy scale which matches the frequency of the driving~\cite{hone2009statistical}.
The further study of this effect is necessary to extend our results.

\begin{acknowledgments}
We thank Dr. Sergio Andraus for carefully reading the manuscript.
This work is supported by KAKENHI Grant No. 25400391.
T.S. acknowledges JSPS for financial support (Grant No. 258794).
T.S. is supported by Advanced Leading Graduate Course for Photon Science (ALPS).
We acknowledge the JSPS Core-to-Core Program ``Non-equilibrium dynamics of soft matter and information.''
\end{acknowledgments}

\end{document}